
\def\sect{\vskip 2mm \centerline}

\def\r{\hangindent=1pc  \noindent}
\def\ref{\hangindent=1pc  \noindent}
\def\cen{\centerline}

\def\v{\vskip 1mm}

\def\endpage{\vfil\break}

\def\kms{km s$^{-1}$}

\def\deg{$^\circ$}

\def\vlsr{$V_{\rm LSR}$}

\def\Vlsr{V_{\rm LSR}}

\def\vrot{$V_{\rm rot}$}
\def\Vrot{V_{\rm rot}}

\def\Vsys{V_{\rm sys}}

\def\tmb{$T_{\rm mb}$}
\def\Tmb{T_{\rm mb}}

\def\Lco{L_{\rm CO}}

\def\Ico{I_{\rm CO}}
\def\ico{$I_{\rm CO}$}

\def\Msun{M_{\odot \hskip-5.2pt \bullet}}

\def\Mhtwo{M_{\rm H_2}}

\def\Deg{^\circ}
\def\deg{$^\circ$}

\def\co{$^{12}$CO($J=1-0$)}

\def\htwo{H$_2$}

\def\ta{$T^*_{\rm A}$}

\def\cc{cm$^{-3}$~}

\def\pa{PASJ}

\def\pal{PASJ Letters}
\def\apj{ApJ}

\def\aj{AJ}
\def\aa{A\&A}

\def\aas{A\&AS}

\def\araa{ARAA}

\def\so{Sofue, Y.}
\def\ha{Handa, T.}
\def\na{Nakai, N.}
\def\fu{Fujimoto, M.}

\def\wi{Wielebinski, R.}
\def\Sofue{Yoshiaki SOFUE}
\def\Nakai{Naomasa NAKAI}

\def\iaut{{\it Institute of Astronomy, University of Tokyo, Mitaka, Tokyo 181}}
\def\nro{{\it Nobeyama Radio Observatory,\footnote*{\rm NRO is a branch of the
National Astronomical Observatory, an inter-university research insittute
operated by the Ministry of Education, Science and Culture.} Minamimaki-mura,
Minamisaku, Nagano 384-13}}


\centerline{\bf CO Observations of Edge-on Galaxies, IV. NGC 4565:}
\centerline{\bf Radial Variation of the \htwo-to-HI Ratio}
\v\v
\cen{\Sofue}
\cen{\iaut}
\cen{and}
\centerline{\Nakai}
\cen{\nro}

\v

\cen{(PASJ: to appear)}

\v
\v
\cen{\bf Abstract}

The edge-on galaxy NGC 4565 has been observed in the  \co-line emission
using the Nobeyama 45-m telescope with an angular resolution of 15$''$.
We obtained a scan along the major axis for $\pm 5' (\pm 15$ kpc) about
the galactic center, and some scans perpendicular to the galactic plane.
The radial density distribution shows a dense molecular gas ring of 5 kpc
radius, which is associated with an HI ring.
The molecular disk comprises two components: an unresolved thin and dense
disk, and a thick disk (or a halo) extending to a height greater than 1.4 kpc.
The position-velocity diagram shows that the general rotation of the entire
galaxy is circular with a flat rotation curve.
However, the radial distribution of molecular gas is asymmetric with  respect
to the galaxy center in the sense that the molecular gas is much richer in
the NW side.
We derived a radial variation of \htwo-to-(HI+ \htwo) density ratio, and
found that  interstellar gas in the central 4 kpc region is almost entirely
\htwo, while HI is dominant  beyond 10 kpc.
A similar HI-vs-\htwo\ behavior was found in the edge-on galaxy NGC 891.

\v

\v\v\v
\sect{\bf I. Introduction}
\v

NGC 4565 is a nearby Sb galaxy of a large angular size with a nearly edge-on
orientation at an inclination angle of 86\deg, and has been extensively
studied by optical observations (de Vaucouleurs 1958; Frankston and Schild
1976; Hamabe et al. 1980; van der Kruit and Searle 1981).
A sharp, straight dark lane runs  along the galactic plane (e.g., Sandage
1961), indicating the presense of a significant amount of interstellar dust,
and therfore a significant amount of molecular gas.
CO line observations using the 7-m telescope at an angular resolution of
100$''$ indicated a centrally peaked molecular gas disk (Richmond and Knapp
1986).
The galaxy has been observed also in radio continuum at various frequencies,
which reveals a thin disk of nonthermal as well as thermal emissions (Hummel
et al. 1984).
Observations of the HI line emission have shown a large disk of interstellar
gas, which is warping in the outermost regions (Sancisi 1976; Rots 1980;
Rupen 1990).

Edge-on galaxies  provide a unique opportunity to investigate the major-axis
distribution of the interstellar gas as well as the distribution perpendicular
to the galactic plane.
However,  since the CO emission in NGC 4565 is relatively weak compared to
other galaxies like NGC 891, there has been no high-resolution observations in
the CO line, and no detailed kinematical characteristics of the molecular gas
has been investigated.

This paper is the fourth of a series describing results of a high-resolution,
high-sensitivity survey of edge-on galaxies in the \co\ line using the
Nobeyama 45-m telescope.
In our Papers I (Sofue et al. 1987) and III (Sofue and Nakai 1992) we
presented results for edge-on galaxy NGC 891, and in Paper II (Sofue et al.
1989) and  in Sofue et al. (1990) we described results for NGC 4631.
In the present paper, we describe results for NGC 4565.

\sect{\bf 2. Observations}
\v

Observations of the \co\ line emission of NGC 4565 were performed in 1992
January and 1992 December using the 45-m telescope of the Nobeyama Radio
Observatory in the course of a CO-line survey of edge-on galaxies.
The parameters for NGC 4565 are shown in table 1.
The systemic LSR velocity is taken to be 1225 \kms according to the HI
helio-centric velocity of 1230 \kms (Sancisi 1976).
We adopt a distance of  10.2 Mpc which was derived by using the Tully-Fisher
relation by Sch{\"o}niger and Sofue (1993).
This value is consistent with the value as suggested by de Vaucouleurs (see
Richmond and Knapp 1986; Rupen 1990).
Hence, the linear scale at the galaxy corresponds to $1'=3.0$ kpc.
{}.

\v\cen{Table 1}\v

Since the observational parameters are described in Paper III, we summarize
them only briefly.
The antenna had a HPBW of $15''$, which corresponds to a linear resolution of
742 pc.
The aperture and  main-beam efficiencies of the telescope were
$\eta_{\rm A}=0.35$ and $\eta_{\rm mb}=0.50$, respectively.
We used an SIS  receiver combined with a 2048-channel acousto-optical
spectrometer of 250 MHz bandwidth corresponding to a velocity coverage of
650 \kms.
After binding up every 32 channels in order to increase the signal-to-noise
ratio, we obtained spectra with a velocity resolution of 10.2 \kms.
The system noise temperatue was 600 to 800 K at observing elevations in the
observations in 1992 January, and 400 to 500 K in December.
We used a multi-on-off switching mode, with which we observed six on-source
positions and two off positions at $5'$ to the east and west in a single
sequence of observing run.

The intensity scale used in the observations was the antenna temparature,
\ta, which is converted to the main-beam brightness temperature by,
\tmb = \ta/$\eta_{\rm mb}=2.0$\ta, and we use \tmb\ in this paper.
The on-source total integration time was about 5 to 10 minutes per each point.
The rms noise of the  resultant spectra at 10 \kms resolution is typically
20 mK in \tmb.
Pointing of the antenna was calibrated every 1 to 1.5 hours using a nearby
SiO maser star at 43 GHz, and the pointing accuracy through the obsevations
was better than $\pm3''$.

We used a coordinate system $(X,Y)$, where  $X$ and $Y$ are defined as the
distances along the major and minor axes from the center position,
respectively ($X$ is positive toward the south-east, and $Y$ is positive
toward the north-east perpendicular to the $X$ axis).
Observations were made at every $15''$ grid interval along the major axis
from $X=-5'$ to $+5'$.
Since the grid interval was equal to the HPBW of the antenna, the data were
under-sampled, so that the final data gave an effective angular resolution of
$\theta=({\rm HPBW}^2+\Delta X^2)^{1/2}=21''$.
Additional observations were made at several positions in the $Y$ direction at
$X=\pm2'$ and $\pm 3'$, by which we confirmed that the CO intensity had a
sharp maximum near the major axis.
We adopted a position angle of the major axis to be 135\deg.5 after Rupen
(1990), and used it in the observations in 1992 December.
However, we used a position angle of 139\deg\ in the observations in 1992
January.
These yielded two sets of data, which referred to slightly different
coordinate systems inclined by a few degrees from each other.
Since the observing points are distributed in a complex way, we conmbined
them to obtain a set of spectra along the major axis at PA=135\deg.5, and we
made use of them to produce a position-velocity diagram.
This diagram practically contains almost all informations which we obtained
from the observations.

\sect{\bf 3. Results}

\sect{\it 3.1. Position-Velocity Diagram}\v

We made use of the composite set of obtained spectra to produce a
 position-velocity (PV) diagram along the major axis at 135\deg.5.
Thereby,  we convolved the data with a gaussian beam with HPBW=$20''$ in the
$X$ and $Y$ direction.
The obatained position-velocity (PV) diagram is shown in figure 1.
The data have been also smoothed in velocity to a velocity resolution of
20 \kms.
This yielded an rms noise of about 12 mK in \tmb\ on the PV map, and the
contours in figure 1 are drawn at every 40 mK \tmb.

\cen{Fig. 1}

The PV diagram provides various informations about kinematics of the galaxy,
 particularly rotation.
A rigid rotation feature is observed at $\vert X \vert < 1'$.
The intensity ridge at  $-1'<X<1'$ can be well defined by an apparently-rigid
rotation feature which has two maxima at $X=\pm 30''$ (1.5 kpc) and
$\Vlsr\sim 1180$ and $ \sim 1300$ \kms, respectively.

In the outer region beyond $X\sim \pm 1'$, the rotation velocity is almost
constant, indicating a flat rotation curve.
By simply tracing the intensity maxima on the PV diagram, we obtain a flat
rotation occurs at \vlsr = 1480 \kms and 980 \kms, which yields a rotation
velocity of $250(\pm10)$ \kms.
This rotation velocity is in a good agreement with  the HI rotation velocity
(255 \kms; Rupen 1990).
{}From these, we can also derive the systemic velocity of $1230 \pm 5 $ \kms,
which is in a good agreement with the HI systemic velocity (1225 \kms; Rupen
1990).
Hence, regardless the significant asymmetry in the gas distribution,
the galaxy is rotating regularly at a constant velocity of 250 \kms
symmetrically about its systemic velocity (1230 \kms).
Such a regular rotation is reasonable in view of the isolation of NGC 4565
from other galaxies.

In addition to these features, a high-velocity component is seen near the
center at $X\sim -30''$ and $ V \sim 1400 - 1500$ \kms, while its opposite
counterpart is lacking or very weak.
This feature could be interpreted as due to a nuclear disk or ring with a
radius of 1.5 kpc, whose distribution is highly asymmetric with respect to
the center.

\sect{\it 3.2. Total Line Profile}\v

Figure 2 shows a ``total line profile" in CO for NGC 4565, which was
obtained by integrating the spectra along the major axis at $-5' < X < 5'$,
or equivalently by integrating the PV diagram in the $X$ direction.
We find a good coincidence with the result obtained by Richmond and Knapp
(1986).
The ``lopsidedness'' of the CO gas distribution, namley much stronger
emission in the positive-velocity side in NW ($X<0'$) than in the
negative-velocity side in SE, is also reproduced in this total profile.

\cen{Fig.2}

In figure 2 we superpose an  HI line profile taken from Rots (1980).
The HI and CO line profiles coincide well with each other.
The total velocity width at 20\% level of peak intensity is measured to be
$540 \pm 10$ \kms, and is about equal to that obtained for the HI emission of
520 to 535 \kms (Rots 1980; Rupen 1990).
This fact supports the argument that the total CO line profiles of galaxies
can be used as the  alternative to the HI Tully-Fisher (1977) relation, as
has been discussed in the case of NGC 891 (Sofue and Nakai 1992) and for other
galaxies (Dickey and Kazes 1992; Sofue 1992; Sch\"oniger and Sofue 1993).

\sect{\it 3.3. Intensity Distribution}\v

The distribution of the integrated  intensity,
$I_{\rm CO} = \int T_{\rm mb}dV$, can be obtained by integrating the
intensity in figure 1 in the direction of velocity, and is shown in  figure 3
as a function of $X$.
The figure indicates a general concentration of CO gas toward the center.
The central strong peak as observed with the $100''$ resolution (Richmond
and Knapp 1986) has been resolved into two peaks at $X\simeq \pm 30''$ in
our observation.
Peaks observed  $X=\pm 3'$  by Richmond and Knapp (1986) coincide with the
sevelar peaks at $X=\pm2-3'$, and finer structures are present.
Also apparent in this figure is the asymmetry of the intensity: it is much
stronger in the NW disk than in SE disk.
It is interesting to note that this asymmetry is consistent with the asymmetry
of dust lane distributions on optical photographs (de Vaucouleurs 1958; Hamabe
et al. 1980), although optical data indicate only foreground absorbing clouds
and not necessarily represent the total dust distribution.

\cen{Fig. 3}

The distribution of radio continuum emission at 20 cm along the major axis is
reproduced from Sukumar and Allen (1991), and indicated by dashed line in
figure 3.
Although we find some global correlation, CO and continuum emissions seem to
be not  well correlated in a few kpc scale length.
In particular, the CO emission seems to avoid the  nuclear continuum source.
Since 20-cm continuum  is dominated by nonthermal emission, it is not adequate
to compare these in the scheme of star-formation efficiency, as we did for
NGC 891 using 6-cm continuum (Sofue and Nakai 1992).

\sect {\it 3.4. Molecular Mass}\v

By integrating all the CO emission in figure 3, we can estimate the total
luminosity in the \co\ emission along the major axis:
The total CO luminosity for the observed region at $-5' < X < 5'$ and
$-7''5<Y<7''5$ (within $\pm 15$ kpc radias and $\pm 370$ pc thickness) is
estimated to be
$\Lco = 4.2 (\pm 0.4)\times10^8$ K \kms pc$^2$.
The error has been estimated as
$\delta L=\sqrt[40({\rm number~of~observed~points})]
\times \delta L({\rm individual~spectrum}),$
where
$ \delta L({\rm individual~spectra})=3 \delta \Tmb({\rm rms})
\delta V \sqrt N$ and $ N=650$ \kms/10 km/s = 65.

If we assume an H$_2$ column density-to-CO intensity conversion factor of
$C=3.6\times10^{20}$ \htwo\ cm$^{-2}$/K \kms (Sanders et al. 1984), the
luminosity can be related to the mass as $\Mhtwo=5.7 \Lco$.
Then,  the total mass of H$_2$ gas within the thin disk of 15 kpc radius is
estimated to  be
$\Mhtwo=2.4 (\pm 0.2)\times10^9\Msun$.
(If we take a smaller conversion factor of
$2.8\times10^{20}$ \htwo\ cm$^{-2}$/K \kms  (Bloemen et al. 1985),
the estimate of the molecular gas mass  should be reduced by a factor of
2.8/3.6=0.78.).
Thus estimated mass is less than the molecular mass obtained by Richmond and
Knapp (1986)($\sim 6 \times 10^9 \Msun$ for the same conversion factor).
The discrepancy may be due to the wider coverage by Richmond and Knapp's
observations along the major axis ($-13' < X <15'$; or $-39<X<45$ kpc) as
well as to their wider beam ($100''=$5 kpc), with which they possibly detected
CO in the halo or a thick-disk (see section 3.8).

\sect{\it 3.5. Radial Density Distribution of the Gas Disk for a Flat Rotation
Curve}\v

The rotation curve of NGC 4565 is flat and the rotation velocity of the galaxy
is about 250 \kms.
Using this fact, we can derive an approximate radial distribution of the
gaseous density for the flat-rotation part of the disk [see Paper III (Sofue
and Nakai 1992) for detail].
According to Paper III, the gas density at radius $R=\vert X \vert$ can be
obtained approximately by
$$ n_{\rm H_2}=C \Ico/l, \eqno(1) $$
where
$$l = 2 \vert X \vert \sqrt{\Vrot^2/V_{\rm min}^2 -1 }. \eqno (2)$$
Here,  $V_{\rm min}$  is the minimum rotation velocity for the integration of
\ico, and \vrot is the rotation velocity, \ico\ is the integrated intensity
around  \vrot, which is obtained by integrating \tmb\ for a range of
$V_{\rm min}<\vert \Vlsr-\Vsys \vert <V_{\rm max}$.
We take $V_{\rm min}=200$  and $V_{\rm max}=280$, which leads to
$ l=1.5 \vert X \vert $.
We note that the spatial density obtained by equatiion (1) indicates a
beam-diluted density, or a density averaged in the $Y$ direction
(perpendicular to the galactic plane) by the beam width (20$''=$ 1 kpc).
Since the molecular gas layer must be much thinner, the true density should be
several times greater than the value indicated here: for example, if we assume
100 pc thickness, the true density can be obtained by multiplying by a factor
10.

Figure 4 shows thus obtained distribution of spatial density of \htwo\
($n_{\rm H_2}$ in H cm$^{-3}$) as a function of radius.
The distribution of $n_{\rm H_2}$ indicates a concentration of \htwo\ gas at
a radius of 5 kpc.
This fact indicates either that the gas distribution is ring-like, or that
 dense spiral arms run at this radius.
Since our observations does not tell about the arm structures, we here simply
interpret this as due to a ring-like (although broad) distribution of
molecular gas.
Figure 4 shows also another strong concentration of molecular gas near the
center, although error and uncertainly of the derived value are large in the
central few kpc.
This central enhancement is due to a possible nuclear disk, which can be seen
in the PV diagram (figure 1).

\cen{Fig. 4}

We apply the same method to an HI intensity distribution in the edge channels
($V_{\rm min}=240$, $V_{\rm max}=280$) around the terminal velocity as
obtained by Rupen (1991) using their 20$''$ resolution data.
In figure 4 we plot thus obtained HI density by a dashed line.
Density indicated here represents also a beam-diluted value, averaged in the
$Y$ direction by the 20$''$ (= 1kpc) HI beam.
The HI density tends to zero inside 4 kpc radius, while it has a sharp ring at
5 kpc radius followed by a broad outer ring-like distribution with the second
peak at 13 kpc radius.
The total HI+H$_2$ density has a clear ring structure at $R\sim 4-5$ kpc.
The \htwo\ molecules are present mainly inside 10 kpc radius, while HI gas has
a broad outskirt reaching as far as 22 kpc radius.
\sect{\it 3.6.  HI-to-\htwo\ Gas Ratio}\v

As shown in figure 4, the HI and \htwo\ gases seem to be present at different
radii, gradually avoiding each other.
Figure 5 plots the mass ratio of \htwo\ gas density to the total (HI+\htwo)
gas density as obtained by using figure 4.
The HI gas is dominant in the outer region beyond 7 kpc, while \htwo\ is
dominant in the inner region.
Particularly for the inner 4 kpc region, the  gas  is almost totally molecular.
The ``exchange'' from HI to \htwo\ appears to occur in coincidence with the
molecular gas ring at 5 kpc radius.

For a comparison, we plot the \htwo/(HI+\htwo) ratio obtained for the edge-on
galaxy NGC 891 by dashed line in figure 5, which has been calculated from the
radial HI and \htwo\ distributions obtained by Sofue and Nakai (1992).
A similar  radial behavior is seen for the exchange from HI to \htwo\ at
around 8-10 kpc radius.
The HI-to-\htwo\ exchange occurs also in this galaxy at 3 kpc, coinciding
with  its 3.4 kpc molecualr ring, inside which the gas is almost perfectly
molecular.

\cen{Fig. 5}

We may be able to interpret this diagram in terms of an evolutionary scenario
of interstellar gases in spiral galaxies as follows.
We may suppose that a primeval galaxy was born as a rotating gaseous disk,
which  comprised primeval HI.
Due to higher gas density in the innermost region of the galaxy, the initial
star formation activity was triggered near the nucleus.
This resulted in an increase in metal abundance in the central region, and,
therefore, in an increase in the circum-stellar formation rate of dust grains.
It is then followed by an increase in interstellar dusts, as represented by
far-IR observations (Wainscoat et al. 1987).
According to the increase in dust grains, interstellar hydrogen is transformed
to \htwo\ molecules through catalystical actions on grain surfaces (e.g.,
Shull and Beckwith 1982).
The action rate is obviously higher near the center according to higher metal
abundance.
Hence, the HI gas in the central region is more rapidly transformed to \htwo,
therefore, to molecular clouds,  than that in the outer region.
Particularly, in the innermost region (inside the molecular rings: $r<4$ kpc
for NGC 4565; 3 kpc for NGC 891), the rate was high enough to transform the
gas to \htwo\ almost entirely, while it was not so high at larger distances
than 10 kpc from the center, and the gas remains to be HI.

Another effect to maintain the higher \htwo-to-HI ratio in the central region
is the increase in the column density of hydrogen molecules and dust grains:
self-shielding by  \htwo\ molecules and absorption by dust prevent UV
radiation from penetrating deep into molecular clouds, and, thus, the
dissociation of \htwo\ into HI is suppressed, particularly, within $r<4$ kpc.

\sect{\it 3.7. The Molecular Gas Ring}\v

Figure 4 indicates an intensity peak at $X\simeq 5$ kpc, and we interpreted
 this peak as due to  a ring of molecular hydrogen gas.
Hereafter, we call this intensity peak the molecular gas ``ring", although
a ring and arms cannot be distinguished from each other since no informatin
about a face-on distribution is available.
Similar concentration of molecular gas at radii of 4 to 5 kpc have been
observed in other edge-on galaxies  so far resolved in CO, and seem to be a
common characteristics to spiral galaxies.
For example, NGC 891 has a ring  with radius of about 3 kpc (Sofue and Nakai
1992), and the Milky Way has the 4-kpc molecular ring (Dame et al. 1987).

Nakai (1992) has suggested that such a molecular gas ring is characteristic
for a barred spiral galaxy.
NGC 4565 has a box- or peanut-shaped bulge (Kormendy and Illingworth 1982),
and such a shape is suggested to have been sustained by a stellar bar (Combes
and Sanders 1981; Combes et al. 1990).
We may thus conjecture that NGC 4565 could be classified as SBb than a normal
Sb.

\sect{\it 3.8. Disk Thickness}\v

In order to clarify if our observations could cover a substantial fraction of
the total emission in the $Y$ direction, or if the molecular gas is extended
in a thicker disk so that we missed some  fraction, we  undertook some scans
perpendicular to the disk plane.
Figure 6 shows an example of spectra taken at $X=2'$ and $3'$ at several
heights ($Y$) perpendicular to the $X$ axis.
The figure indicates that, although the CO intensity decreases rapidly with
the height, weak emission can be seen above $Y\simeq\pm15''$.
Figure 7 shows a plot of integrated CO intensity across the galactic plane
(as a function of $Y$) at $X=+2'$.
This figure indicates that a significant emission is detected beyond
$Y\sim \pm 15''$.
This fact shows that the molecular gas distribution comprises two components:
a thin and dense disk, whose thickness is smaller than the beam size
(0.7 kpc), and a thick diffuse disk (or a halo), which extends at least up to
$Y\sim\pm 30-45''$ with a scale height greater than 1.4 kpc.
The emission from the extended component is comparable to that from the
galactic plane covered by our $15''$ beam.

Using the intensity distribution along the major axis obtained by Richmond and
Knapp (1986) with their $100''$ beam, we estimated  ``total'' \htwo\ mass
within $-5'<X<5'$ to be $\sim 5 \times 10^9\Msun$, which is a mass contained
within an area of $10'\times100''=(30$ kpc $\times 5$ kpc).
This mass is significantly greater than the presently derived mass of
$2.4 (\pm 0.2) \times10^9\Msun$ along the galactic plane within
$10'\times15''$ (30 kpc $\times 0.74$ kpc) in section 3.4.
This discrepancy may imply that a substantial fraction of \htwo\ mass (the
rest $\sim 2.6\times 10^9 \Msun$) is distributed beyond the thin disk that
was covered by the present beam of $15''$ HPBW.
This is consistent with the extended emission as shown in figure 7.
The dynamical mass of the galaxy within  radius $r= 15$ kpc can be estimated
to be  $M_{\rm dyn}\sim 2.0 \times 10^{11}\Msun$ from the rotation velocity of
250 \kms.
Hence, if we include the thick disk component, approximately 2.5\% of the
total mass is taken by the molecular gas within the same radius, while only
1\% is taken by the thin molecular disk of thickness less than 0.7 kpc
($15''$).

Alternatively, the apparently extended component beyond $Y\sim\pm 15''$ could
be explained by a tilted disk by 4\deg\ from the line of sight (de Vaucouleurs
1958), even if the disk thickness is small enough.
For example, emission from the disk plane at 10 and 20 kpc distance from the
major axis (node) is observed at $Y\sim \pm 0.7$ and 1.4 kpc, respectively.
In this case,  the radial velocity of the emission should vary with the
distance and therefore with  $Y$: the larger is $\vert Y \vert$, the closer
is the radial velocity to the systemic velocity.
However, a careful inspection of figure 6 reveals no such variation of the
radial velocity with $Y$ at least up to $Y\sim \pm 30-45''$, but it indicates
a corotation of the extended component with the disk gas.
Hence, we may conclude that the extended emission in figure 7, and probably
the larger amount of molecular gas observed by larger beam of $100''$ by
Richmond and Knapp (1986), is due to an extended molecular gas halo or thick
disk.

\sect{\bf 4. Summary}\v

We mapped the major axis of the edge-on galaxy NGC 4565 in the \co\ line
emission using the Nobeyama 45-m telescope at an angular resolutin of $15''$
(=742 pc).
The results are summarized as follows:

(a) The position-velocity diagram indicates a flat rotation of the galaxy
disk, and its rotation characteristic is symmetric with respect to the
galactic center as well as to the systemic velocity. The rotation velocity of
the flat part is 250 \kms.
Although the general rotation is symmetric,
the molecular gas distribution is  asymmetric with respect to the galactic
center.

(b) The molecular mass involved in the thin disk within 15 kpc radius is
estimated to be $2.4(\pm 0.2) \times10^9\Msun$, which takes only 1\% of the
total (dynamical) mass within the same radius.
In addition to the thin and dense disk, a comparable amount of molecular gas is
 present in a thick disk or halo with a scale height greater than 1.4 kpc.

(c) The intensity and radial density distributions along the galactic plane
shows a central condensation of molecular gas  with rigid rotation, and a
ring-like distribution at 5 kpc radius.
 This 5-kpc molecular ring is associated with an HI ring, followed by an outer
 extended HI disk.

(d) The \htwo-to-(HI+\htwo) mass ratio increases toward the center.
Interstellar gas in the innermost region at $R<4$ kpc is almost entirely
molecular, while it is HI dominant outside 10 kpc radius, and the HI to \htwo\
exchange occurs at around 4 kpc just at the inner edge of the 5-kpc ring.
The diagram could be interpreted as to represent a histroy of the overall
conversion of HI to \htwo\ in a disk galaxy, and should be  useful for
discussion of an evolution of interstellar gas in a disk galaxy.

\v
The authors thank the staff of NRO for their help during the observations.

\sect{\bf References} \v

\r Bloemen, J.B.G.M., Strong, A. W., Cohen, R. S., Dame, T. M., Grabelsky,
D. A., Hermsen, W., Lebrun, F., Mayer-Hasselwander, H. A., and Thaddeus, P.
1985, \aa, {\bf 154}, 25.

\r Combes, F., Debbasch, F., Friedli, D., and Pfenniger, D., 1990, \aa,
{\bf 233}, 82.

\r Combes, F., and Sanders, R. H. 1981, \aa, {\bf 96}, 164.

\r{Dame, T. M., Ungerechts, H., Cohen, R. S., de Geus, E. J., Grenier, I. A.,
May, J., Murphy, D. C., Nyman, L. -A, and Thaddeus, P.  1987, \apj, {\bf 322},
706. }

\r de Vaucouleurs, G. 1958, \apj, 127, 487.

\r Dickey, J., and Kazes, I. 1992, \apj, in press.

\r Frankston, M., and Schild. R. 1976, \aj, 81, 500.

\r Hamabe, M., Kodaira, K., Okamura, S., and Takase, B. 1980, \pa, 32, 197.

\r \ha, \na, \so,  M.Hayashi, and \fu 1990, \pa, {\bf 42}, 1.

\r Hummel, E., Sancisi, R., and Ekers, R. D. 1984, \aa, {\bf 133}, 1.

\r{Huntley, J. M., Sanders, R. H., and Roberts, W. W.,  1978, \apj, {\bf 221},
521.}

\r \na 1992, \pal, 44, in press.

\r Kormendy, J., and Illingworth, G. 1982, \apj, {\bf 256}, 460.

\r Richmond, M. W.,  and Knapp, G. R. 1986, \aj, {\bf 91}, 517.

\r Rots, A. 1980, \aas, {\bf 41}, 189.

\r Rupen, M. P., 1990, \aj, {\bf 102}, 48.

\r Sancisi, R. 1976, \aa, {\bf 53}, 159.

\r Sandage, A. R. 1961, {\it The Hubble Atlas of Galaxies} (Carnegie
Institution, Washington), p. 25

\r Sanders, D. B., Solomon, P. M., and Scoville, N. Z. 1984, \apj, {\bf 276},
 182.

\r Sch\"oniger, F., and \so\ 1993, \aa, in press.

\r Shull, J. M., and Beckwith, S. 1982, \araa, {\bf 20}, 163.

\r \so, 1992, \pal, 44, L231.

\r \so, \ha, and \na   1989, \pa, {\bf 41}, 937 (Paper II).

\r \so, \ha, Golla, G., and \wi  1990, \pa, {\bf 42}, 745.

\r \so, and \na 1992, \pa, submitted (Paper III).

\r \so, \na, and \ha  1987, \pa, {\bf 39}, 47 (Paper I).

\r S$\phi$rensen, S. -A., Matsuda, T., and Fujimoto, M. 1976, {\it Astrophys.
Sp. Sci. }, {\bf 43}, 491.

\r Sukumar, S., and Allen, R. J. 1991, \apj, {\bf 382}, 100.

\r Tully, B., and Fisher, J. R. 1977, \aa, {\bf 54}, 661.

\r van der Kruit, P. C., and Searle, L. 1981, \aa, 95, 105.

\r Wainscoat, R. J., de Jong, T., and Wesselius, P. R. 1987, \aa, {\bf 181},
 225.

\endpage

\settabs 7 \columns
\noindent{Table 1. Parameters for NGC 4565}
\v

\hrule\v

\+The center position ($X=0'',~Y=0''$)&& &&NED$^\dagger$ \cr

\+ ~~~  R.A.$_{1950}$  & \dotfill& $12^{\rm h} 33^{\rm m} 51^{\rm s}.8$ \cr

\+ ~~~  Decl.$_{1950}$ & \dotfill& $26\Deg 15' 50.''0$ \cr

\+ PA of the major axis & & 135\deg.5 & &Rupen (1990) \cr

\+  Systemic velocity (\vlsr)  \dotfill& & 1230 ($\pm 5$) \kms && Present CO
result \cr

\+ \vrot &  \dotfill& 250 ($\pm 5$) \kms & &Present CO result \cr

\+ Distance &	\dotfill& 10.2 Mpc &&Sch{\"o}niger and Sofue (1993)  \cr

\v
\hrule
\v
$\dagger$ NASA/IPAC Extragalactic Database (NED) (1991) (operated by Jet
Propulsion Laboratory, California Institute of Technology under contract by
NASA).

\endpage

\noindent {\bf Figure Captions} \v

Fig. 1: A Position-velocity $(X-V)$ diagram along the major axis of NGC 4565.
The resolution in the map is $\Delta X \times \Delta V = 20'' \times 20$ \kms,
and the effective rms noise of the map is about 12 mK in \tmb.
Contours are drawn at every 40 mK in \tmb\ starting at 40 mK, and the peak
intensity in the map is 420 mK \tmb.
The cross marks the galaxy center and the systemic velocity.

Fig. 2: A total line profile of the \co\ emission along the major axis.
Velocity resolution in this diagram is 10 \kms.
An HI total line profile is shown by the dashed line (Rots 1980).

Fig. 3: \ico\ distribution along the major axis $X$ (full line), and intensity
of 20-cm continuum emission (dashed line: Sukumar and Allen 1991).
The resolution in the $X$ direction is $20''$.

Fig. 4: Beam-diluted spatial densities of molecular and neutral hydrogen gases
(in H \cc), averaged in the $Y$-direction by a  beam size of  20$''$ (1 kpc).

Fig. 5: The mass ratio of \htwo\ to total (HI+\htwo) gas density as a function
of the galactocentric distance.

Fig. 6: \co\ spectra at various heights ($Y$) from the galactic plane at
$X=2'$ and $3'$.

Fig. 7: CO intensity distribution perpendicular to the galactic plane at
$X=+2'$ as a function of $Y$.
The distribution comprises two components: a thin and dense disk, and an
extended thick disk or a halo.
Dashed line represents a gaussian beam with HPBW=15$''$.

\bye